# Quantum two-way time transfer over a 103 km urban fiber

Huibo Hong, Runai Quan, Xiao Xiang, Yuting Liu, Tao Liu, Mingtao Cao, Ruifang Dong, and Shougang Zhang

*Abstract*—As a new approach to realizing high-precision time synchronization between remote time scales, quantum two-way time transfer via laboratory fiber link has shown significant enhancement of the transfer stability to several tens of femtoseconds. To verify its great potential in practical systems, the field test in long-haul installed fiber optic infrastructure is required to be demonstrated. In this paper, we implement the two-way quantum time transfer over a 103 km urban fiber link. A time transfer stability of 3.67 ps at 10 s and 0.28 ps at 40000 s has been achieved, despite the large attenuation of 38 dB leading to fewer than 40 correlated events per second. This achievement marks the first successful step of quantum two-way time transfer in the task of high-precision long-distance field transfer systems.

## I. INTRODUCTION

NOWADAYS, time synchronization has become a ubiquitous task in national well-being and development, which has been applied from precise navigation [1] to finance [2] and modern communication [3], [4], from critical infrastructures [5]–[9] to quantum information processing [10]. Among different time transfer methods [11]–[15], the two-way time transfer (TWTT) is widely used as its achievable time stability can be immune from the actual path delay fluctuation. Further investigation also shows that the two-way approach enables the detection of man-in-the-middle (MITM) delay attacks, making it a requisite for secure time synchronization [16]. Based on the TWTT method, the satellite link has achieved the time stability better than 100 ps over a day of averaging in terms of time deviation (TDEV) [17]. Due to the low loss, high reliability, and high stability of optical fibers, time transfer over fibers offers potentially superior performance against satellite-based counterparts [14], [15]. Over a 50 km dark urban fiber link, the TWTT achieves a minimum stability of 0.6 ps at an average time of 1000 s [18]. To further explore the advantage of the two-way time transfer, quantum enhanced two-way time transfer (Q-TWTT) was first proposed in 2017 [19]. According to quantum theory, the strong temporal correlation in energy-time entangled photons ensures the narrow correlation peak in their arrival times at distributed parties [20]. Therefore, quantum enhancement beyond the possibilities of classical physics is highly expected via Q-TWTT and a proof-of-principle experiment was subsequently implemented over a 20-km coiled fiber link. The result shows a TDEV of 922 fs at 5 s and 45 fs at 40960 s [21]. Meanwhile, the verified feature of no correlation between the measured clock difference and the propagation distance has shown that the Q-TWTT setup is robust against symmetric channel delay attacks [22], [23]. Shortly afterward, the experimental demonstration was extended to a 50 km coiled fiber with the TDEV stability reaching 2.6 ps at 7 s, and a minimum of 54.6 fs at 57300 s [24]. In comparison with its classical counterpart [25], the Q-TWTT result has shown significant improvement in the time transfer performance, especially in the long-term averaging time[26], [27]. Its verification in metropolitan urban fiber links is urgently expected to pave the way for the field applications of Q-TWTT in long-haul fiber-optic timing systems.

In this paper, we report the implementation of the Q-TWTT over a 103 km urban fiber link between the central campus of National time service center (NTSC-CC) in Lintong District and its Hangtiancheng campus (NTSC-HTC) in Chang'an District. Despite the rather few correlated detection events as low as 40 per second due to the significant loss up to 38 dB, the time stability achieved 3.67 ps at averaging time of 10 s and 0.28 ps at 40000 s. Maintaining the detected idler photon counts basically the same, the time stability over a 103 km coiled fiber was also investigated, which gave the results of 2.37 ps at 10 s and 0.28 ps at 40000 s. The overlap between the two TDEV curves beyond 20000 s shows the presence of non-reciprocal delay variation related to fiber length, which limits the long-term stability. This experimental demonstration fully exhibits the great potential of quantum two-way time transfer in field applications for synchronization stability far below 1 picosecond.

This work was supported by the National Natural Science Foundation of China (Grant Nos. 12033007, 12203058, 61801458, 61875205, 12074309, 91836301), the Strategic Priority Research Program of CAS (Grant No. XDC07020200), the Youth Innovation Promotion Association, CAS (Grant No. 2021408, 2022413, 2023003), the Western Young Scholar Project of CAS (Grant Nos. XAB2019B17 and XAB2019B15), the China Postdoctoral Science Foundation(2022M723174).
(Corresponding author: Ruifang Dong, Shougang Zhang) Huibo Hong and Runai Quan contributed to the work equally and should be regarded as co-first authors.

The authors are with Key Laboratory of Time and Frequency Primary Standards, National Time Service Center, Chinese Academy of Sciences, Xi'an, 710600, China and School of Astronomy and Space Science, University of Chinese Academy of Sciences, Beijing, 100049, China (e-mail: honghuibo@ntsc.ac.cn; quanrunai@ntsc.ac.cn; xiangxiao@ntsc.ac.cn; liuyuting@ntsc.ac.cn; taoliu@ntsc.ac.cn; mingtaocao@ntsc.ac.cn; dongruifang@ntsc.ac.cn; szhang@ntsc.ac.cn).

## II. EXPERIMENTAL SETUP

The experimental setup for the Q-TWTT over a 103 km-long urban fiber path is sketched in Fig.1. The two parallelly installed urban fiber links (Link 1 and Link 2) from NTSC-CC to NTSC-HTC have a single-trip length of about 51.5 km. By joining the ends of Link 1 and Link 2 at NTSC-CC, the 103 km fiber link was formed. To evaluate the quantum enhanced time transfer performance, the self-reference method was adopted. That is, the two sites to be synchronized (site A and B) were co-placed in the lab at NTSC-HTC. The self-developed all-fiber telecom-band energy-time entangled biphoton source with a center wavelength of 1560 nm [28] was utilized to address the timing offset [29]. In principle, two separate energy-time entangled biphoton sources should be utilized for the Q-TWTT experiment. Due to limited entangled sources for the multiple experiments being conducted in parallel, we split the source into two for the experimental implementation. The achievable time stability has been verified equivalent to that using two separate sources with high spectral consistency. The generated polarization-orthogonal photon pairs were spatially departed and output from the ports of the source equipment. Subsequently, the signal photons were split into two portions by a 50/50 fiber beam splitter (FBS1), with each one followed by a fiber optical circulator (OC1 & OC2). By respectively connecting OC1 and OC2 (shown by indigo lines) with the ends of Link 1 and Link 2, the bidirectional transmission paths of the two 'signal beams' were constructed. Meanwhile, the idler photons were also split by another 50/50 fiber beam splitter (FBS2). The two outputs of FBS2 were then respectively connected via the optical circulators OC3 and OC4 to the two ends of a dispersion compensating fiber module (DCFM, BG-DCM-100KM), whose nominal group delay dispersion (GDD) is equivalent to that of the 100 km fiber except with opposite sign. Thus, the two bidirectionally transmitted 'idler beams' nonlocally compensated the dispersion experienced by the 'signal beams' [30].

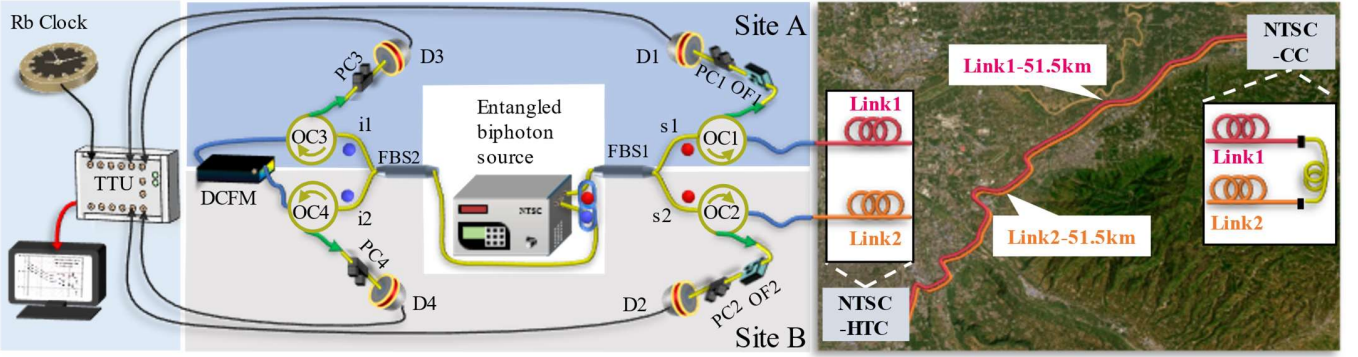

**Fig. 1.** Experimental diagram of the quantum two-way time transfer over the 103 km urban fiber optic link. For simplified implementation, sites A and B are co-located and share an energy-time-entangled photon pair source. The signal photons are split into two portions and counter-propagated through the urban fiber loop formed by Links 1 and 2, and then detected by the superconducting nanowire single-photon detectors D1 and D2. While the idler photons are detected by D3 and D4 after being transmitted through the dispersion-compensated fiber module (DCFM). By being referenced to the Rb clock, the time-tagging unit (TTU) is used to record the photon arrival times at D1-D4. OC: optical circulator; OF: optical filter; PC: fiber-based polarization controller.

After each port 3 (the outputs are denoted by green lines) of OC1-OC4, a superconducting nanowire single-photon detector (SNSPD, respectively marked as D1-D4) with 80 % quantum efficiency and about 50 ps RMS (Root mean square) timing jitter was set to detect the transmitted signal photons and idler photons. In front of each SNSPD, a fiber-based polarization controller (PC) was inserted to ensure the initial polarization of the injected photons compensate for the polarization selectivity of the SNSPD. To suppress the significant excess noise contribution from the urban fiber link, fiber bandpass filters centered at 1560 with a passband width of 6.5 nm in FWHM (Full width at half maximum) (OF1 & OF2) were inserted in front of D1 and D2. The arrival times were then stamped by a commercial time-tagger instrument (TTU, Time tagger ultra, Swabian Instruments) that was referenced to a common Rb clock (Ultra, Rock electronic LTD). The detected photon arrival events by D1, D2, D3 and D4 were respectively tagged as $t_1$, $t_2$, $t_3$, $t_4$. From the recorded arrival times, the temporal correlation measurements were accomplished to extract the time differences $t_1 - t_3$ and $t_2 - t_4$ [29]. According to the two-way principle, the time offset $t_0$ was given by $((t_1 - t_3) - (t_2 - t_4))/2$. By monitoring the variation of $t_0$ with the measurement time, the uncertainty and stability performances of the Q-TWTT system were evaluated.

## III. RESULTS AND ANALYSIS

For our urban fiber link, about 88 % of them are buried under the ground while the rest are hanged on. In the experiment, extremely large and unwanted accidental counts were found that not only submerges the signal photons but also saturates the detection of the SNSPDs (~600 kcps). To evaluate the noise count contribution in the urban fiber link, we first inserted a programmable optical filter (Waveshaper 1000B, POF) in front of D1. By setting the pass-through bandwidth (BW) of the POF to be 0.5 nm ± 0.08 nm and scanning their center wavelength from 1525 nm-1600 nm with a step of 0.5 nm, the spectral distribution of the unwanted noise counts in the fiber links was analyzed by the detection of D1. As shown by the red up-triangles in Fig. 2(a), three peaks were located around 1530 nm, 1544 nm, and 1551 nm. In addition, there was an almost uniformly distributed background noise present throughout the entire scanning range, which contributed a 'white' count rate of approximately 800 cps for each scanned spectral window. Upon inspection, the high noise count peak around 1530 nm and the "white" background noise originated from the Erbium-Doped Fiber Amplifiers (EDFAs) installed at the end of some adjacent optical fibers in the optical cable, which were 200 meters of fiber distance apart from the SNSPD D1. While the peaks around 1544 nm and 1551 nm were due to the crosstalk interferences from the parallelly transmitted optical carriers for other time/frequency transfer and internet services in the adjacent fibers (more detailed discussions can be found in the Appendix).

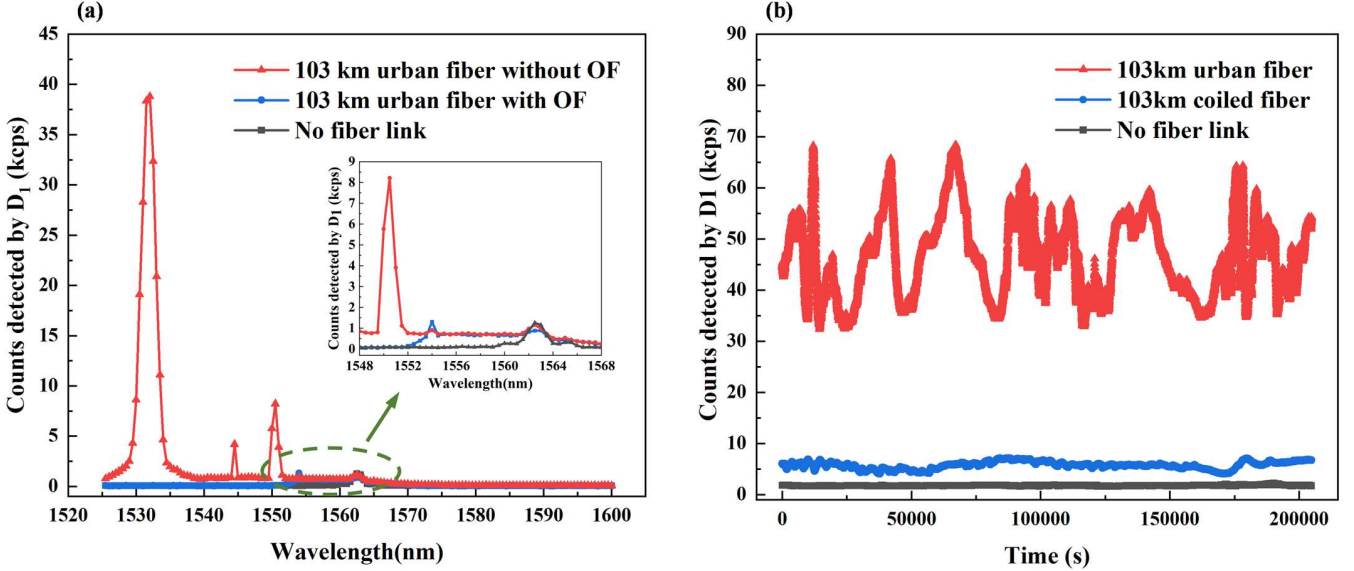

**Fig. 2.** (a) Spectral distribution of D1-detected signal photons in three transmission cases: 103 km urban fiber without optical filter (red triangles), 103 km urban fiber with optical filter (blue circles) and no urban fiber (black squares). For perspective, the inset shows the three cases zoomed in the 1548 nm – 1568 nm region. (b) The observed signal photon fluctuation behaviors by D1 in the cases of transmitting through the 103 km urban fiber (red triangles), the 103 km coiled fiber (blue circles), and no fiber link (black squares).

Fortunately, the spectral center of the utilized photon pair source is at 1561 nm with a FWHM width of 1.84 nm [28], as shown by the black squares in the inset of Fig. 2(a), these noise contributions can be effectively filtered out from the actual signal photons by applying fiber bandpass filters (OF1 & OF2) with the passband at 1560 nm $\pm$ 6.5 nm in front of D1 and D2. After applying OF1, the detected signal photon count rate of D1 as a function of the POF center wavelength is shown by the blue circles in the inset of Fig. 2(a). After experiencing the significant link loss induced by the 103 km urban fiber (~29 dB) as well as the insertion loss by the OCs and OF (~2 dB), the signal photon count rate detected by D1 as a function of measuring time is then plotted in Fig. 2(b) by red up-triangles. A violent fluctuation of the detected photon counts from 30 kcps to 70 kcps can be seen. For comparison, the signal photon count rate detected by D1 in the cases of after transmitting through a 103 km coiled fiber link and a no-fiber link (built by short-cutting OC1 and OC2 with a 1-m-long fiber and optical attenuator, and removing the DCFM) respectively are also plotted in Fig. 2(b) with blue circles and black squares. One can observe that, although the photon count fluctuation is reduced magnificently, there is still nontrivial fluctuation in the case of 103 km coiled fiber transmission compared with the case of no fiber link. Through analysis, such fluctuations should be related to the polarization sensitivity of the SNSPD for photon detection. The photons transmitted through the optical fiber link may undergo increased polarization variations in proportional to the fiber length as well as the ambient interference, resulting in significant fluctuation in the photon detection.

The detection of the idler photons after passing through the relevant OCs and DCFM, which contribute a loss of ~6 dB, were maintained at a count rate of 400 kcps. After taking into account the quantum efficiency of the SNSPDs (~1 dB per SNSPD), a total loss of about 38 dB should contribute to the temporal coincidence measurement between the signal and idler photons. Under the harsh loss and violent fluctuation over the urban fiber link, the Q-TWTT experiment was implemented.

Due to the significant loss, the achieved coincidence events involved in the one-way temporal correlation measurements for $t_1 - t_3$ and $t_2 - t_4$ were as few as 40 per second. In order to give a reliable temporal coincidence distribution of the photon pairs, the measurement time for each run was set as 10 s. Based on Gaussian fitting of the histograms for $t_1 - t_3$ and $t_2 - t_4$, showed widths of about 188 ps and 175 ps in FWHM (see Table 1) implied that most of the dispersions experienced by the signal photons in the 103 km fiber have been effectively cancelled by the DCFM inserted in the idler arms. From Gaussian fitting of the histograms, the measured peak values of $t_1 - t_3$, $t_2 - t_4$, and $t_0$ within the interval of 10 s can be extracted. Fig. 3(a) depicts the variations of $t_1 - t_3$ (blue squares), and $t_0$ (black line) over the time duration for more than 2 days ($2\times10^5$ s). As can be seen, the one-way time difference ($t_1 - t_3$) underwent a quite significant fluctuation reaching a peak-to-peak value up to 12.9 ns, which showed an apparent periodic behavior in close relevance to the diurnal variation of the ambient temperature over the fiber link. By assuming an average temperature change of 2 degrees Celsius over the urban fiber route, which is reasonable for a buried fiber, a similar value of 14 ns for the single-path time delay variation over the 103 km fiber was estimated. Whereas under the two-way configuration, the fluctuation of the extracted time offset $t_0$ is also shown in Fig. 3 (a) by black squares, in which no apparent dependence on the ambient temperature change over the fiber link is observed. The standard deviation (SD) is calculated to be only 4.0 ps, whose comparability with the predicted result of 3.7 ps according to the theoretical simulation [31] further reflects that the Q-TWTT system has a nice bidirectional symmetry. For comparison, the measurement of $t_0$ over time in the cases over the 103 km coiled fiber link and no-fiber link were also performed while maintaining the detected idler photons count rate at the same level of 400 kcps. The results are plotted in Fig. 3 (b) and (c) respectively. With the 103 km coiled fiber link, the measured coincidence distributions for $t_1 - t_3$ and $t_2 - t_4$ showed widths of about 198 ps and 182 ps in FWHM. The SD of $t_0$ was achieved as 2.9 ps and agreed well with the theoretical prediction of 2.5 ps. With no fiber link, the coincidence widths for $t_1 - t_3$ and $t_2 - t_4$ were measured as 116 ps and 86 ps, which represented the jitter contributions from the SNSPDs and set the limits for the achievable minimum coincidence widths in each path (see Supplementary information for more discussions). The corresponding SD of $t_0$ was achieved as 1.6 ps in experiment and 1.5 ps in theory. From the gap of measured coincidence widths between the cases with and without the 103 km fiber in the link, incomplete dispersion cancellation can be inferred. According to our previous study [32], due to the nonzero spectral width of the SPDC pump, the minimum temporal correlation widths after the 103 km fiber transmission in bidirectional paths would be limited to 139 ps and 115 ps under optimized NDC, inferring that the SD of the time offset can be optimized to 2.6 ps under the same coincidence events.

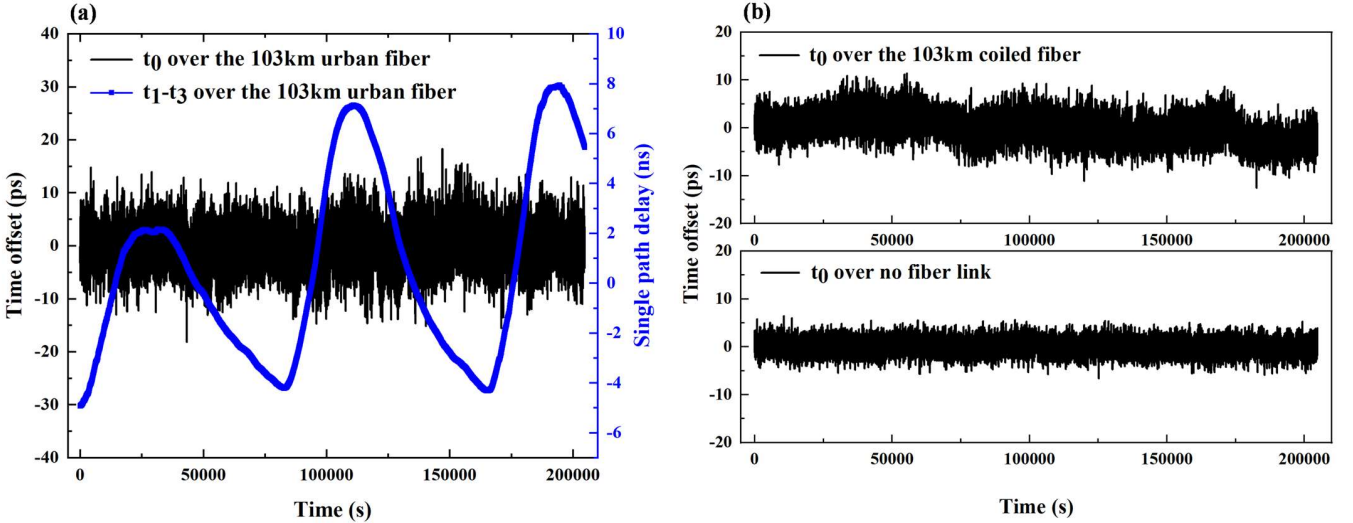

**Fig. 3.** The measured variation plots over time. (a) $t_1 - t_3$ (blue squares) and $t_0$ (black dots) over the 103 km urban fiber link; (b) $t_0$ over the 103 km coiled fiber link; (c) $t_0$ over no fiber link.

TABLE 1 NO FIBER LINK, 103KM COILED FIBER, 103KM URBAN FIBER FOR THE CASE OF THEORETICAL AND MEASURED SD

|  | $t_1 - t_3$ coincidence width (ps) | $t_1 - t_3$ coincidence events (in 10 s) | $t_2 - t_4$ coincidence width (ps) | $t_2 - t_4$ coincidence events (in 10 s) | Theoretical SD of $t_0$ (ps) | Measured SD of $t_0$ (ps) |
|---|---|---|---|---|---|---|
| No fiber link | 116 | 846 | 86 | 796 | 1.5 | 1.6 |
| 103 km coiled fiber | 198 | 1058 | 182 | 974 | 2.5 | 2.9 |
| 103 km urban fiber | 188 | 436 | 175 | 412 | 3.7 | 4.0 |

The uncertainty of the Q-TWTT over the 103 km urban fiber link was investigated as well. Based on the expression shown in Ref. [33], the variance of the extracted time offset can be categorized into the following three terms:

$$\Delta^2 t_0 = \Delta^2 \left[ \frac{(t_1 - t_3) - (t_2 - t_4)}{2} \right]$$
$$+ \frac{1}{4}(\Delta^2(t_1 - t_3)_{cal} + \Delta^2(t_2 - t_4)_{cal})$$
$$+ \frac{1}{4}\Delta^2(\tau_{FAB} - \tau_{FBA}) + \frac{1}{4}\Delta^2(\tau_{DCF,A} - \tau_{DCF,B}). \quad (1)$$

The first term represents the uncertainty of the measured time differences based on the two-way time transfer system. From the long period of time difference measurements conducted over the 103 km urban fiber, it was evaluated as 4.0 ps. The second term denotes the calibration uncertainty of the overall setup for distributing, receiving, detecting the photon pairs, and recording their arrival times. The contributed elements include the optical fiber-based BSs, OCs, OFs, and the SNSPDs, TTU, etc. By shortcutting the fiber link and removing the DCFM, this uncertainty was also evaluated from the time difference measurements of $t_1 - t_3$ and $t_2 - t_4$, which yielded the FWHM widths of 86 ps and 116 ps respectively. As the calibration should be made under the same condition of optical attenuation with the case over the 103 km urban fiber link, the coincidence events given by the last row of Table 1 were utilized for evaluating the calibration uncertainty. Based on the theoretical model given in [31], this calibration uncertainty can be analyzed as 2.1 ps. The third term denotes the non-reciprocal time delay variation in the fiber link. In practical applications, two entangled biphoton sources should be used. Therefore, slight spectral inconsistency between the two sources would introduce a non-reciprocal time delay in the fiber link that should be calibrated. According to Ref. [34], it can be evaluated by $\tau_{FAB} - \tau_{FBA} = LD\Delta\lambda/2$, where $L$ is the fiber length, $D \sim 17$ ps/nm/km is the dispersion of the single mode fiber, and $\Delta\lambda$ is the center wavelength difference of the signal photons of the two entangled sources. However, due to environmental variation, such as temperature, experienced by the fiber and the entangled sources, as well as the measurement errors in determining parameters like the dispersion coefficient, fiber length and optical wavelengths, the non-reciprocal delay variation should be considered even after calibration. Assuming a 2 degrees Celsius variation in fiber temperature and a center wavelength difference of 0.02 nm, as stipulated in [24], the non-reciprocal time delay variation due to temperature fluctuations in the fiber link can be regarded as negligible. On the other hand, there should be an uncertainty of determining the center wavelength of signal photons due to certain thermal sensitivity and measurement error. During a 39-hour period of monitoring the signal photons' center wavelength, a standard deviation of 3 pm was resulted, leading to a type-A uncertainty of 3.7 ps. For the temperature-induced fluctuation of the signal photons' center wavelength, both the thermal sensitivity of the signal photons' wavelength and the precision of the temperature controller for the PPLN waveguide in the experiment [28][28] were applied. According to our previous study [35], the thermal sensitivity of the signal photons' wavelength was measured to be 0.4 nm/°C. As the nominal precision of the temperature controller is ±0.05°C, the temperature-induced fluctuation of the signal photons' center wavelength in SD can be estimated to be 7 pm, thus resulting a type-B uncertainty contribution of 8.3 ps. Combining the two contributions, the total uncertainty is 9.1 ps. Furthermore, the non-reciprocal time delay variation resulting from the polarization mode dispersion (PMD) effect and the Sagnac effect should be taken into account [33]. Using the typical coefficient of PMD, $\sim 0.05$ ps/$\sqrt{\text{km}}$, the PMD-dependent uncertainty for the 103 km fiber was evaluated as 0.25 ps. The uncertainty associated with the Sagnac effect correction was estimated to be 2.6 ps by taking a typical value of 0.05 ps/km [36]. Similarly, the uncertainty contribution from the fourth term was evaluated based on the determination accuracy of the idler photons' center wavelength, resulting in an equivalent value of 9.1 ps. All in all, the combined uncertainty for the Q-TWTT system over the 103 km urban fiber was estimated as 13.9 ps (see Table 2).

TABLE 2. UNCERTAINTY BUDGET FOR Q-TWTT OVER 103 KM URBAN FIBER LINK

| Uncertainty source | Uncertainty contribution (ps) | Uncertainty type |
|---|---|---|
| Time difference measurement | 4.0 | A |
| System calibration | 2.1 | A |
| Non-reciprocal delay from the fiber link | 9.1 | A&B |
| Non-reciprocal delay from the DCFM | 9.1 | A&B |
| PMD effect | 0.25 | B |
| Sagnac effect | 2.6 | B |
| Combined standard uncertainty | 13.9 | |

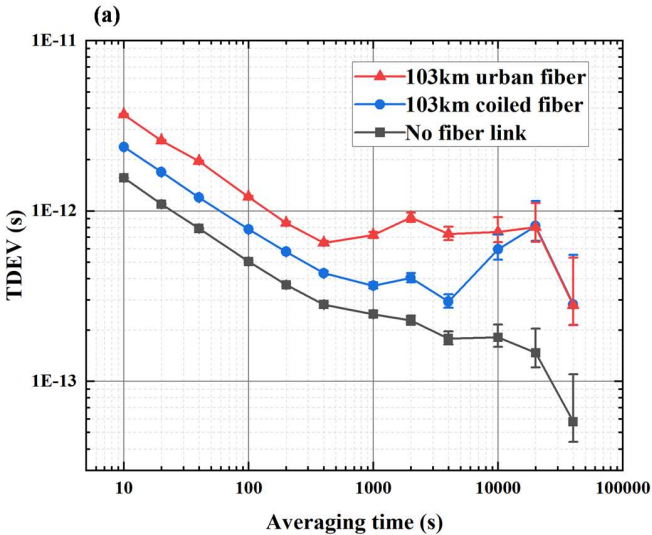 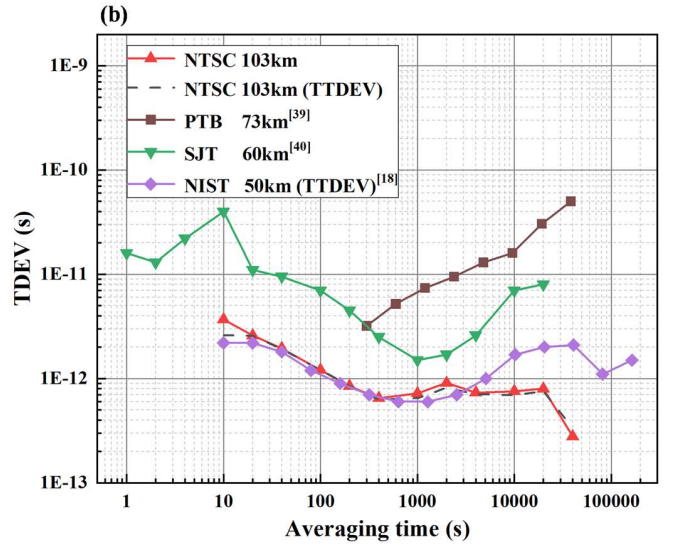

**Fig. 4.** (a) TDEV of 103 km urban fiber (red up-triangles), 103 km coiled fiber (blue circles), and without fiber link (black squares). (b) the time transfer stability performance comparisons of Q-TWTT and TWTT.

The corresponding time stability performances were evaluated in terms of time deviation (TDEV) over the entire 2.5 days of measurement period and the results are plotted in Fig. 4(a). With no fiber link (as depicted by black squares), the TDEV result gives the lower limit for the achievable system stability. It reaches a value of 1.56 ps at averaging time of 10 s and a minimum of 58 fs at 40,000 s, which is consistent with that in [21], [27]. With the 103 km urban fiber link (as shown by red up-triangles), the TDEV is 3.67 ps at 10 s and reaches a minimum of 0.28 ps at 40,000 s. With the 103 km coiled fiber link (as shown by blue solid circles), the TDEV is 2.37 ps at 10 s and has the same minimum of 0.28 ps at 40,000 s. The three TDEV curves all follow a descending slope of $\tau^{-1/2}$ before the averaging time of 400 s. The gap of about 1ps from each other is mainly due to the differences of the achieved coincidence events and recovered temporal coincidence widths in the three cases. Beyond the averaging time of 400 s, the TDEV curve for the case over no fiber link presents a slower downward trend than the $\tau^{-1/2}$ slope starting from 1000 s to the averaging time of 20000 s. It should be attributed to the effect of periodical temperature variation in the air-conditioned laboratory on the experimental system [37]. In the case over the 103 km coiled fiber link, as the periodical temperature variation also has an impact on the fiber link, bulges are observed between 1000 s and 20000 s. Since the 103 km urban fiber link experienced a much more dramatic temperature variation, the corresponding TDEV curve ceases to fall down when the averaging time is beyond 400 s and the apparent bulge maintains until the averaging time arrives at 20000 s. Due to changes in environmental temperature and pressure along the fiber route, the optical birefringence of the fiber experiences a random time-varying behavior, e.g., the polarization-mode dispersion (PMD) effect, which cannot be eliminated by the bidirectional propagation. Such non-reciprocal delay variation has been found associated with the fiber length [38] and limits the long-term stability. From Fig. 4 (a), the TDEV of 103 km urban fiber (red up-triangles) shows an apparent deterioration between 400 s and 20000 s. As mentioned above, about 88 % of our urban fibers are installed underground while the rest are hanged on. For the 12 % hanged fibers, rapid variation over minutes may occur due to significant day-night temperature changes. While for the installed fibers, the drift of the optical polarization is normally on the order of hours to days. The variance property due to the PMD has a good consistency with the TDEV of 103 km urban fiber. Note should be taken that, there is PMD in the coiled fiber as well. Therefore, the TDEV curve of the 103 km coiled fiber shows similar behavior after the averaging time of 20000 s. However, the TDEV curves in both cases over the 103 km urban fiber link and over the 103 km coiled fiber link are consistently better than 1ps after an averaging time of 150 s. The long-term TDEV results reach a similar minimum of 0.28 ps. To visually showcase the advancement of Q-TWTT compared with the classical TWTT approach. Fig. 4(b) gives some of reported TWTT results over the intercity-distance field fibers ranging from 50 km to 73 km [18], [39], [40]. One can further confirm that the Q-TWTT has superior symmetry in bidirectional transmission.

## V. SUMMARY AND DISCUSSION

In summary, the two-way quantum time transfer over a 103 km-long urban fiber loop has been successfully implemented with an evaluated uncertainty of 13.9 ps and a minimum time stability of 0.28 ps achieved. The superiority of Q-TWTT over the classical TWTT, especially in the long-term time stability, reveals its great potential in the metro and inter-city optical fiber time synchronization systems. In the experiment, as the utilized DCFM did not realize a perfect NDC of the 103 km-long fiber, a further improvement of the short-term stability to 2.62 ps at 10 s averaging could be expected by optimizing the NDC. Meanwhile, the DCFM is about 12 km-long and would also introduce ambient fluctuation to the system [21]. By replacing it to the fiber Bragg grating (FBG) module, which is normally only 10-m long, the long-term time stability will also be improved.

## APPENDIX

In the 51.5 km optical cable between NTSC-HTC and NTSC-CC, there are 22 optical fibers. We used 2 of them for our Q-TWTT experiment. The other optical fibers are also in use: 2 for transferring the standard time and frequency signal (corresponding optical carriers are at 1543.73 nm and 1542.94 nm), 2 for implementing the 9 GHz microwave frequency transfer (corresponding optical carriers are at 1551.72nm and 1550.92nm), 2 for implementing the optical frequency transfer (at 1550.12 nm), and the rest fibers are for providing the internet service (corresponding optical carriers are at 1550 nm). Therefore, the crosstalk interferences from adjacent fibers contribute unwanted accidental counts that not only submerge the signal photons but also saturate the detection of the SNSPDs. The noise photon spectra were measured and shown in Fig. 2 (a). As can be seen, three peaks located around 1530 nm, 1544 nm and 1551 nm are observed. The peaks around 1544 nm and 1551 nm should be due to the crosstalk induced by the above optical carriers. While the highest and broad peak at 1530 nm is owing to the working EDFAs installed at the end of the other fibers in the engine room at NTSC-HTC, which is 200 m apart from our SNSPDs. To clearly demonstrate the crosstalk interference due to the EDFA, we have executed the noise photon spectra measurement with only one EDFA involved. By turning on and off the electric power of the EDFA, the noise photon spectra are shown in Fig. A1 by red dotted line and black square line respectively. One can see that, the working EDFA not only induces significant noise counts at the 1530 nm band but also leads to certain amplification of the crosstalk at 1544 nm and 1550 nm. Meanwhile, it is the source of lifting the background counts within the scanned spectral window of 0.5 nm from ~115 cps to ~295 cps. Note should be taken that, the dark count rate of the SNSPDs is only about 60 cps and the background counts of ~115 cps per 0.5nm is contributed by the fluorescence from liquid crystal in the wave shaper. Furthermore, we investigated the dependence of this EDFA-induced crosstalk interference on the distance between the EDFA and the SNSPD D1. The results are shown in Fig. A2, the degradation of the noise photon counts with the increase of the fiber length can be clearly seen, which further proves the noise contribution from the EDFA.

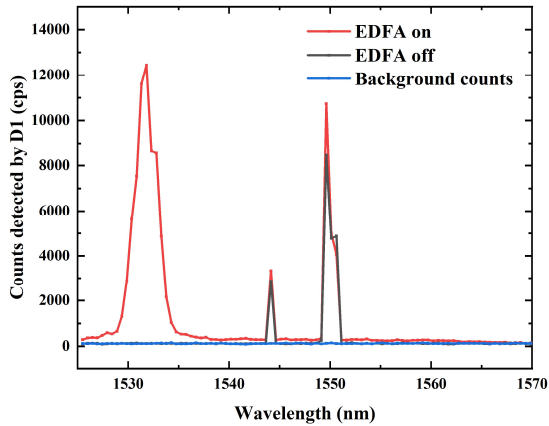

**Fig. A1.** D1-detected noise spectral distributions when the single EDFA is powered on (red dotted line) and off (black square line).

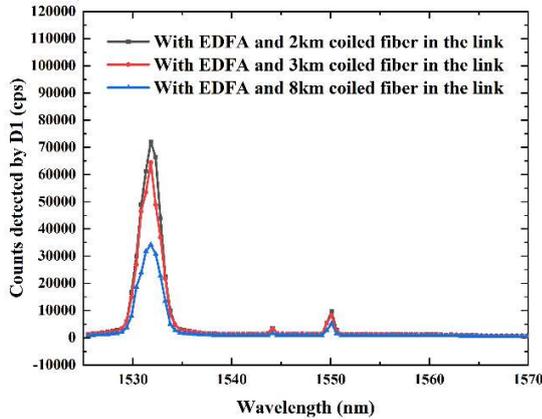

**Fig. A2.** D1-detected noise spectral distributions with the inserted fiber length between the EDFA and D1 being set as 2 km (black), 3 km (red), and 8 km (blue) respectively.

SUPPLEMENTARY INFORMATION

In our previously reported experiment of the 50 km Q-TWTT [24], commercial event timers (ETs, Eventech Ltd, A033-ET/USB) were used to record the arrival times of the detected photons. Due to the sampling rate limitation of the event timers, the photon count rate of each SNSPD was set around 20 kcps. Under such photon count rate condition, the FWHM timing jitter about 60 ps was contributed from the SNSPDs. In this experiment, to combat the much larger loss induced by the 103 km-long urban fiber link, the commercial time tagger instrument (TTU, Time tagger ultra, Swabian Instruments) was applied, which could support a maximum detection count rate of 70 Mcps (Mega-counts per second). Therefore, the photon count rates of the SNSPD D3 and D4 in this experiment were set to be 400 kcps. Such optimization can effectively improve the measurement efficiency of the Q-TWTT system. However, as the increment of the photon count rate detected by the SNSPDs will also lead to the broadening of the timing jitter for the SNSPDs [41]. The tradeoff between the photon count rates and the timing jitter is hence an important issue for the optimization of the short-term time stability.

4